\documentstyle[11pt]{article}

\begin{document}
\begin{titlepage}
  \begin{flushright}
    KUNS-1569\\[-1mm]
    hep-ph/9904433
  \end{flushright}

  \vspace*{5ex}
  \begin{center}
    {\Large\bf
      On fermion mass hierarchy with extra dimensions}
    \vspace{5ex}

    Koichi Yoshioka\footnote{E-mail address:
      yoshioka@gauge.scphys.kyoto-u.ac.jp}
    \vspace{1ex}

    {\it Department of Physics, Kyoto University, Kyoto 606-8502,
      Japan}
  \end{center}
  \vspace*{8ex}

  \begin{abstract}
    Recently, various phenomenological implications of the existence
    of extra space-time dimensions have been investigated. In this
    letter, we construct a model with realistic fermion mass hierarchy
    with (large) extra dimensions beyond the usual four dimensions. In
    this model, it is assumed that some matter fields live in the
    bulk and the others are confined on our four dimensional wall. It
    can naturally reproduce the quark and lepton mass hierarchy and
    mixing angles without any symmetry arguments. We also discuss some
    possibilities of obtaining suitable neutrino masses and mixings
    for the solar and atmospheric neutrino problems.
  \end{abstract}
\end{titlepage}

To understand the hierarchical structure of the fermion masses and
mixings is one of the most important problem in the particle
physics. In the Standard Model, all the Yukawa couplings are
completely free and they are nothing but phenomenological input
parameters. To reproduce fermion masses or at least to relate one mass
(Yukawa coupling) to others the informations on some dynamics at
higher scale are needed beyond the Standard Model. One important
approach is to consider grand unified theories (GUT), in which
frameworks the quarks and leptons are also unified into the same
multiplets of unified gauge groups and their Yukawa couplings are
related each other at high energy. This is a very beautiful approach, 
but still there are large ambiguities in determining the Yukawa
couplings of three generations. In addition, the GUT gauge symmetries
generally relate quarks and leptons and so their mixing angles. This
fact simply induces a contradictory small lepton mixing, and then we
have to deal with rather complex models to reproduce realistic
low-energy values of couplings. In $SU(5)$ GUTs, however, the
left-handed (charged-) leptons are associated with only the
right-handed down quarks, not with the left-handed quarks, resulting
in possible large lepton mixing without having unrealistic CKM
angles. This interesting feature has been recently discussed and used
to construct models with large lepton mixing \cite{mixing}.

Recently, the notable approach to models beyond the Standard Model is
proposed \cite{LED,tev-gut,hierarchy}. This approach invokes the
existence of large extra dimensions in addition to our four
dimensions. This interesting possibility has been used to argue many
problems in the present particle physics and 
astrophysics \cite{astro}, and its various effects on future
experiments are also intensively discussed \cite{exp}. With these
large extra dimensions, we can have many renewed insights into those
have been discussed in traditional four dimensional frameworks; the
gauge hierarchy problem \cite{hierarchy}, gauge coupling 
unification \cite{tev-gut,unify}, fermion mass 
hierarchy \cite{tev-gut,abel,ext-mass}, neutrino 
physics \cite{neutrino}, proton decay, and so on.

In this letter we demonstrate that in a GUT framework with (large)
extra dimensions, the observed fermion mass hierarchy and mixing
angles which include the large lepton mixing can be naturally
reproduced. As for the size of the extra dimensions $R$, there has
been many works in which $R^{-1}$ is much lower than the
four-dimensional Planck scale; $R \sim O({\rm mm})$ or  
$O({\rm TeV}^{-1})$. In such frameworks, many new interesting
possibilities occur as mentioned above. On the other hand, for the
most part of this paper we consider the compactification of the
extra dimensions occurs around the usual GUT scale $\sim 10^{16}$ GeV
and the fundamental scale $M_*$ is taken as the order 
of $10^{18}$ GeV\@. Taking the compactification scale $R^{-1}$ near
the GUT scale also has several attractive features from
phenomenological points of view. Below the GUT scale, we have the
Minimal Supersymmetric Standard Model (MSSM) and so the traditional
gauge coupling unification by interpolating the experimental 
data \cite{lep2} can be preserved. In addition, the Kaluza-Klein (KK)
threshold corrections at the GUT scale may improve a little larger
prediction of the low-energy $SU(3)$ gauge coupling in 
the MSSM \cite{su3}. The proton stability can be also guaranteed more
trivially than in the low-scale unification scenarios, and the
supersymmetry breaking problems (the flavor problem, etc.)\ and many
other problems could be solved by considering the compactification
just below the Planck scale \cite{randall}.

To obtain hierarchy among Yukawa couplings, we need have the
origins of suppression of these couplings. In the traditional four
dimensional models, these small factors originate from some
hierarchies among couplings themselves, a small ratio to the
fundamental scale, the radiative corrections, and so on. It implies
that we have to invoke some specific dynamics such as flavor
symmetries in order to have hierarchical couplings. On the other hand,
in models with extra spacetime dimensions, we generally have some
possible suppression mechanisms for couplings. First possibility is a
volume factor accompanied with each bulk field in integrating out the
extra dimensions \cite{astro,neutrino}. In case of one extra
dimension, this volume factor $\epsilon$ associated with its mode
expansion becomes
\begin{eqnarray}
  \epsilon \,\simeq\, \frac{1}{(M_*R)^{1/2}}\,,
  \label{epsilon}
\end{eqnarray}
to have a canonically normalized four dimensional field. For the
fields which can propagate in $n$ extra dimensions, this volume factor
becomes $1/(M_*R)^{n/2}$. Therefore, when there are different fields
propagating in different numbers of dimensions, different volume
factors are attached to their couplings in four dimensions. If the
size of the extra dimensions is relatively large, this factor can
produce hierarchy among couplings. Secondly, the other possible effect
is that some couplings have the power-law evolution behaviors in
their renormalization group equations (RGE) due to the existence of
infinite KK towers in the bulk \cite{lr,tev-gut,abel}. In other
words, this is due to the fact that the higher dimensional couplings
have non-zero classical mass dimensions and then under the classical
scaling, each coupling suffers the power-law corrections. This large
effect can vary the low-energy values of coupling constants
considerably and in some cases we can obtain hierarchical structures 
among these couplings in the form of $1/(M_*R)^x$ as well as the first
possibility. Though other interesting mechanisms which depend on some
specific dynamics have been proposed so far \cite{ext-mass}, in this
paper we focus on the first possibility. We will also mention that the
second approach, namely, the RGE effects can be used alternatively to
obtain the same results as in the models constructed by the first
approach.

Let us explain our model. We consider a supersymmetric $SU(5)$ GUT
model with 3 generations ${\bf 10}$ and ${\bf 5}^*$ representation
matters and 3 singlets as the right-handed neutrinos. Among these
matter fields, we assume that the first generation ${\bf 10}_1$ can
propagate in two extra dimensions (the fifth and sixth) and the second
generation ${\bf 10}_2$, one extra (the fifth) dimension. The $SU(5)$
gauge fields are also in the bulk and all other fields are confined on
the four dimensional wall (our world). This configuration, in which
the different fields feel the different numbers of extra dimensions,
can be actually realized in the Type I string 
theory \cite{tev-gut}. The extra dimensions are compactified on tori
with a common radius $R$ and its scale is supposed to be on the order
of the GUT scale ($R^{-1} \simeq 10^{16}$ GeV). In this situation,
we have suppression factors in the interaction terms concerning 
the ${\bf 10}_{1,2}$ fields due to the above mentioned volume 
factor; they are $\epsilon^2$ accompanied with ${\bf 10}_1$ and
$\epsilon^1$ with ${\bf 10}_2$. Since each Yukawa coupling is given
by the superpotential ${\bf 10}_i {\bf 10}_j H_u$ for the up-type
quarks, ${\bf 10}_i {\bf 5}^*_j H_d$ for the down-type quarks and the
charged leptons, and ${\bf 5}^*_i {\bf 1}_j H_u$ for the neutrinos,
the order of magnitude of the quark and lepton Yukawa couplings just
below the GUT scale ($\simeq$ the compactification scale) become as
follows;
\begin{eqnarray}
  \label{mass-up}
  Y_u &\simeq& \left(
    \begin{array}{ccc}
      \epsilon^4 & \epsilon^3 & \epsilon^2 \\
      \epsilon^3 & \epsilon^2 & \epsilon   \\
      \epsilon^2 & \epsilon   &  1
    \end{array} \right),\\
  \label{mass-down}
  Y_d &\simeq& \left(
    \begin{array}{ccc}
      \epsilon^2 & \epsilon^2 & \epsilon^2 \\
      \epsilon   &  \epsilon  &  \epsilon  \\
      1    &   1  &  1
    \end{array} \right),\\
  Y_e &\simeq& \left(
    \begin{array}{ccc}
      \epsilon^2 & \epsilon & 1 \\
      \epsilon^2 & \epsilon & 1 \\
      \epsilon^2 & \epsilon & 1
    \end{array} \right),\\
  \label{mass-neu}
  Y_\nu &\simeq& \left(
    \begin{array}{ccc}
      1  &  1  &  1  \\
      1  &  1  &  1  \\
      1  &  1  &  1
    \end{array} \right).
\end{eqnarray}
These matrices are denoted in the basis of $L_iY_{x_{ij}}R_j$ form
where $L(R)$ represents the left(right)-handed fermions. It should be
noted that since the gauge fields are assumed to live in the bulk, the
four-dimensional gauge coupling constant is only normalized 
as $g\to g/\sqrt{V}$ ($V$: the volume of bulk) independently of the
properties of the matter fields. Therefore, we have no hierarchical
strength of gauge coupling to different matter fields. In deriving the
above matrix forms we assume that all the Yukawa couplings are of the
same order of magnitude in higher dimensional (higher scale) theories,
which can be naturally expected. Of course, there also exist other
possibilities of taking specific forms of Yukawa couplings which may
be restricted by the group theoretical assumptions, some flavor
symmetries, etc. Such choices could improve the inaccurate relations
in detailed analyses but we do not mention this possibility in this
letter.

In the Yukawa matrices (\ref{mass-up})--(\ref{mass-neu}), we have an
ambiguity of $O(1)$ coefficient in each element. Therefore each matrix
is generally supposed to have non-zero determinant. By diagonalizing
the above matrices, we obtain the following fermion mass hierarchies;
\begin{eqnarray}
  m_u : m_c : m_t &\simeq& \epsilon^4 : \epsilon^2 : 1, \\
  m_d : m_s : m_b &\simeq& \epsilon^2 : \epsilon : 1, \\
  m_e : m_\mu : m_\tau &\simeq& \epsilon^2 : \epsilon : 1.
\end{eqnarray}
Now the compactification scale is supposed to be around the GUT scale
and then the suppression factor becomes $\epsilon \sim 0.1$ when
taking $M_*$ as the typical string scale $\sim 10^{18}$ GeV (see 
Eq.\ (\ref{epsilon})). This value is roughly consistent with the
experimental mass eigenvalues of quarks and leptons in the 
large $\tan \beta$ case. Note that the larger hierarchy in the 
up-quark sector than in the down-quark and charged-lepton sectors is
remarkably accomplished. Exactly speaking, the first generation
down-quark and charged-lepton masses do not agree well with the
experimental values. This point will be improved later, in that case
more matter fields live in higher dimensions beyond the four.

Another important result of the above matrices is the mixing angles of 
quarks and leptons. It can be seen from the Yukawa 
matrices (\ref{mass-up}) and (\ref{mass-down}) that the quark mixing
angles are small. This is due to the fact that both the up- and
down-type left-handed quarks belong to the same ${\bf 10}$ multiplets
which have the hierarchical property in the present model. On the
other hand, the left-handed charged leptons are in the ${\bf 5}^*_i$
and there is no special structure among them. This fact indicates 
that in contrast to the quark sector, the lepton mixing angles are
large which is actually required to solve the observed anomalous
results in the neutrino experiments. As for the neutrino side, we do
not have any conditions on the right-handed neutrinos fields, that is,
we deal with them equally. In this situation, it is natural to 
naively assume an approximate permutation symmetry and then no
hierarchical structure in the right-handed Majorana mass matrix. Then,
the structure of the left-handed Majorana mass matrix can be
determined by the properties of ${\bf 5}_i^*$ only as long as the
seesaw mechanism \cite{seesaw} is applied in order to have small
neutrino masses. Consequently we expect the large mixing between
generations in the neutrino side. After all, we can have the large
mixing angles in the lepton sector unless the accidental cancellation
between the charged-lepton and neutrino mixing matrices occurs.

Here we comment on another possibility of having a hierarchical
factor from the existence of extra dimensions, that is, the power-law
effects from the renormalization group equations. For example, we
consider a slightly different $SU(5)$ model than before. In this
model, all three generation matters are now confined on the wall and
instead of it, there is a new gauge singlet field $\theta$ which lives
in the bulk. In addition, we assume a $U(1)$ flavor symmetry under
which the $\theta$ field has charge $-1$ and the ${\bf 10}_1$ and
${\bf 10}_2$ also have charge $+2$ and $+1$, respectively, and all
other fields have zero charge. In this model, hierarchical 
parameters are provided by the power-law evolutions of Yukawa
couplings due to the presence of the KK excitations associated with
the $\theta$ fields \cite{tev-gut}. For example, the above $U(1)$
symmetry permits a superpotential 
term $y\, \theta^4 {\bf 10}_1 {\bf 10}_1 H\,$ where $y$ denotes a
dimensionful `Yukawa' coupling. This term induces the couplings 
of KK modes of $\theta$ to the matter field ${\bf 10}_1$ on the
wall. With these couplings, summing up the one-loop graphs which
include the KK modes of $\theta$ induces a correction which is
proportional to $y^2\,(M_*R)^{4n}$ to the anomalous dimension 
of ${\bf 10}_1$. Here $n$ denotes the number of extra dimensions
that the $\theta$ feels, and the contribution of Higgs KK modes is
neglected because it is a common contribution to all the anomalous
dimensions of matter fields.\footnote{The Higgs anomalous dimensions
  can also be neglected if we assume that the Higgs fields feel some
  numbers of   extra dimensions.} \ By rescaling $y$ to the
dimensionless Yukawa coupling $Y_{u_{11}}$, it can be easily seen that
the correction to $Y_{u_{11}}$ becomes of 
order $(M_*R)^{-(2n+4)}$. This is just the low-energy value 
of $Y_{u_{11}}$ itself, assuming large Yukawa couplings at high-energy
scale (the quasi fixed-point behavior \cite{quasi}). By similar
analyses of the other Yukawa couplings we obtain the same forms of
matrices as (\ref{mass-up})--(\ref{mass-neu}), now by replacing the
suppression factor $\epsilon \simeq (M_*R)^{-1/2}$ with 
$\epsilon' \equiv \epsilon^{(n+2)/2}$. In case $n=1$, for example,
the requirement that $\epsilon' \sim 0.1$ implies 
$M_*R \sim 30$. Note that the hierarchical values of Yukawa
couplings are realized as the quasi fixed-point values. This fact
indicates that each Yukawa coupling is very large at high energy and
in addition its low-energy value is determined independently of the
high-energy informations. With these reasons a volume factor
associated with the singlet $\theta$ is not the dominant feature in
this approach. In this way, we can reproduce the same results as that
in the previous scenario. This translation is actually possible for
any matrix form. In addition, in the present model there is no
standard gauge non-singlet fields in the bulk. This may be preferable
from some phenomenological points of view (proton decay etc.).

The mass matrix forms of the Eqs.\ (\ref{mass-up})--(\ref{mass-neu})
have been already discussed before in the context of the $SO(10)$
grand unified theory \cite{so10} and the supersymmetric composite
model \cite{composite}. In these models, the hierarchical parameters
can be obtained by setting the other parameters or the compositeness
scales of technicolor gauge groups. This fact usually requires some
specific situations (the special couplings or gauge groups, etc.) in
order to have hierarchical structure. The present mechanism of large
volume factors from the extra dimensions, however, can always be
applied in any (GUT) models, and has wide possibilities of
solving other phenomenological problems.

In this way, we can have the roughly consistent predictions to the
low-energy values of masses and mixings. In the present models, it is
essential to accomplish the natural hierarchy that 
the ${\bf 10}$ representation matters live in the bulk or have
non-zero charges under some flavor symmetry. However, this situation
has nothing to do with the neutrino mass matrix. The realistic mass
difference for solving the solar neutrino problem, by the matter
enhanced (MSW) oscillation solution \cite{MSW} or the vacuum
oscillation scenario \cite{vacuum}, can be obtained by an accidental
cancellation among the couplings or by introducing the other small
parameters. In the following we discuss this point while considering
the several possible improvements of the previous simple scenario.

First, we consider the case in which some more matter fields also feel
the extra dimensions. In particular, we assume that the first
generation 5-plet, ${\bf 5}^*_1$, feels one (the fifth) extra
dimension besides the ${\bf 10}_{1,2}$. This induces further
suppression factors and then we have the following Yukawa matrices;
\begin{eqnarray}
  Y_d &\simeq& \left(
    \begin{array}{ccc}
      \epsilon^3 & \epsilon^2 & \epsilon^2 \\
      \epsilon^2 &  \epsilon  &  \epsilon  \\
      \epsilon   &   1  &  1
    \end{array} \right),\\
  Y_e &\simeq& \left(
    \begin{array}{ccc}
      \epsilon^3 & \epsilon^2 & \epsilon  \\
      \epsilon^2 & \epsilon   &  1  \\
      \epsilon^2 & \epsilon   &  1
    \end{array} \right).
  \label{new-e}
\end{eqnarray}
The up-quark and neutrino Yukawa couplings are not affected by this
change. On the other hand, the left-handed neutrino Majorana mass
matrix takes the following form, assuming no constraint (no hierarchy)
among the right-handed neutrino masses as mentioned before,
\begin{eqnarray}
  m_\nu^L &\simeq& \left(
    \begin{array}{ccc}
      \epsilon^2 & \epsilon  & \epsilon  \\
      \epsilon   &  1  &  1  \\
      \epsilon   &  1  &  1
    \end{array} \right)\frac{v^2}{M_N},
  \label{new-nu}
\end{eqnarray}
where $M_N$ is a typical right-handed neutrino Majorana mass and $v$ is
the vacuum expectation value of the electroweak Higgs doublet. The
mass eigenvalues of quarks and leptons now become,
\begin{eqnarray}
  m_u : m_c : m_t &\simeq& \epsilon^4 : \epsilon^2 : 1, \\
  m_d : m_s : m_b &\simeq& \epsilon^3 : \epsilon : 1, \\
  m_e : m_\mu : m_\tau &\simeq& \epsilon^3 : \epsilon : 1.
\end{eqnarray}
This hierarchical structure is more realistic than before, namely, the
down-quark and charged-lepton masses of the first generation are 
improved. Exactly speaking, however, the $O(1)$ coefficients (the
Clebsch-Gordon coefficients) may be needed to reproduce the actual
low-energy values in the down-quark and charged-lepton sector. Note
that these mass matrix structures can also be obtained in the model
with the power-law running effects as described before, in which 
the ${\bf 5}_1^*$ field is now assumed to have charge $+1$ under 
the $U(1)$ flavor symmetry.

As for the neutrino masses and mixing, the mixing between the first
and second generations is small $\sim O(\epsilon)$ and the 
mixing between the second and third generations is expected to be
large $\sim O(1)$ in the neutrino and charged-lepton sectors. At first
sight, the latter result seems to require rather fine-tunings among
the Yukawa couplings. However, the matrix $Y_eY_e^\dagger$, which the
left-handed mixing matrix for the charged lepton can diagonalize, has
the same form as $m_\nu^L$ (\ref{new-nu}). Since these two matrices
are given by the Yukawa couplings squared, we can compare the degrees
of tuning in diagonalizing these matrices. In the matrix form 
of $Y_e$ (\ref{new-e}) which is often discussed in the literatures it
is usually assumed that the muon mass eigenvalue is reproduced by the
tuning of couplings. The necessary order of fine-tuning 
is $(m_\mu/m_\tau)^2 \sim 10^{-3}$ from the experimental values. On
the other hand, in the $m_\nu^L$, we need a tuning of the order 
of $(\Delta m^2_{\rm sol}/\Delta m^2_{\rm atm})^{1/2} \sim 10^{-1}$
to have the experimentally required mass-squared differences for the
neutrino anomalies. Therefore, we can more naturally realize a mass
hierarchy in the neutrino sector than that in the charged-lepton
sector which is often assumed. In this situation, the masses of the
light neutrinos from the above matrix (\ref{new-nu}) become,
\begin{eqnarray}
  m_{\nu_1} : m_{\nu_2} : m_{\nu_3} \;\simeq\; 
  \epsilon^2 : (\Delta m^2_{\rm sol}/\Delta m^2_{\rm atm})^{1/2} : 1 
  \;\;\simeq\; \epsilon^2 : \epsilon : 1.
\end{eqnarray}
For $\epsilon = (M_*R)^{-1/2} \sim 0.1$, this hierarchy shows the
required mass eigenvalues for $M_N \sim 10^{14-15}$ GeV\@. This 
value of $M_N$ is relatively larger than the typical intermediate
scale favored by cosmology and astrophysics. But, there is a
simple solution. That is, if the ${\bf 5}^*$ fields feel more extra
dimensions, $M_N$ can become lower. More interestingly, this choice
also reduces the down-quark and charge-lepton mass eigenvalues and
could realize the small $\tan \beta$ case. After all, we can see that
the solar neutrino problem is solved by the small angle MSW solution
between the first and second generations and the atmospheric neutrino
anomaly, by the large mixing between the second and third
generations. Other oscillation solutions are also possible. As
mentioned in Ref.\ \cite{composite}, if we assume that the $2\times 2$
sub-matrix of $m_\nu^L$ has rank 1, it leads a bi-maximal solution in
which the solar neutrino problem is solved by the large angle MSW
solution instead of before. The vacuum oscillation scenario is not
readily achieved because the large hierarchy between the second and
third generations in the neutrino sector leads too small down-quark
and electron masses. This may be improved if there are other origins
of hierarchy in the right-handed Majorana mass matrix independently of
the properties of right-handed neutrinos which, for example, comes
from other singlet fields. 

We finally discuss the possible existence of larger size of extra
dimensions, $\sim O({\rm mm})$ and/or $\sim O({\rm TeV}^{-1})$
scale. In the models discussed so far, we have derived the light
neutrino mass matrix by taking an assumption of the heavy Majorana
masses and utilizing the seesaw mechanism. However, since the
right-handed neutrinos are the standard gauge singlets they can
propagate in these large extra dimensions without any
problems. This can produce different and interesting phenomenology to
neutrino physics. Consider an additional assumption than the previous
$SU(5)$ models that the right-handed neutrinos ${\bf 1}_i$ are also
bulk fields. In this case, a large volume factor with these 
right-handed bulk neutrinos leads a large suppression of the
neutrino Dirac masses and one may no longer need heavy fields to have
small masses. Then it has been shown that the $O({\rm mm})$ size of
extra dimensions are just the required order of magnitude for
explaining the neutrino problems \cite{neutrino}. The detailed
analyses in this framework show that the solar neutrino problem can be
solved by the small angle MSW solution (the oscillation to the bulk
neutrinos) but the large mixing angle for the atmospheric neutrino
anomaly is excluded by the astrophysical limits \cite{dvali}. However,
in the present $SU(5)$ model, since the large mixing angle between the
second and third generations can be obtained from the charged-lepton
side we can expect to have a natural solution for both neutrino
problems without heavy fields.

In this scenario, the gravity is also extended to higher
dimensions. This implies that since the realm of quantum gravity is
now open at rather lower scale $M_*$, the simple unification
description at high-energy scale is no longer relevant. Then we
suppose that the compactification scale $R^{-1}$ of the extra
dimensions which the standard non-singlet fields can feel is as low as
order TeV, namely, a low-scale unification scenario \cite{tev-gut}. 
We also assume that the number of the large extra 
dimensions ($r \sim O({\rm mm})$) is two\footnote{This is also needed
  to have three chiral families in a construction from the Type I
  string theory \cite{shiu}.} in order to avoid the gravitational
experimental limits and to obtain interesting neutrino phenomenology
as mentioned above. Then it can be easily seen from the 
relation $M_{\rm pl}^2 = M_*^{\delta+2}R^{\delta-2}r^2$ that the
remaining $\delta$ extra dimensions are consistently around the scale
of $M_*$, which can be as low as TeV\@. We can use these 
freedom (the $\delta$ extra TeV$^{-1}$-sized dimensions) to have
fermion mass hierarchy among the generations as described
before. After all, we consider a model in which the matters
corresponding to ${\bf 10}_{1,2},$ etc.\ live in some of these
$R$-sized extra dimensions as previously and the right-handed
neutrinos and gravity live in larger $r$-sized dimensions. Since in
this model the whole $SU(5)$ multiplets always together feel extra 
dimensions, the gauge coupling unification surely occurs in a similar
way to Ref.\ \cite{tev-gut}. As for the Yukawa couplings, the
hierarchical structure is obtained in the same way as described
before. The suppression volume factor is now numerically evaluated
from the gauge coupling unification scale $\Lambda\, (\simeq M_*)$ as 
\begin{eqnarray}
  \epsilon \,\simeq\, \frac{1}{(\Lambda R)^{n'/2}} \;\sim\;
  \frac{1}{\sqrt{20}}\,.
  \label{gut-ep}
\end{eqnarray}
Note that in the above formula, $n'$ denotes the maximal number of
extra dimensions in which the standard gauge non-singlet fields
live. This implies that $\epsilon$ in Eq.\ (\ref{gut-ep}) is the
minimum suppression factor accompanied with such bulk fields and the
others are larger than $\epsilon$. In addition, one more serious
problem arises in this low-scale unification scenario; the
proton-decay amplitude must be suppressed. In case no matter multiplet
can propagate through more than four dimensions, several mechanisms to
avoid the proton decay have been proposed; the (discrete) 
symmetries \cite{tev-gut,shiu,discrete}, some specific 
dynamics \cite{ext-mass}, and so on. Now, if only the ${\bf 10}$
fields live in the bulk, some discrete symmetry could forbid the
dimension five baryon number violating operators. However, naively the
dimension more than six operators from the K\"ahler potentials are not 
forbidden and we would have to rely on other model-dependent
mechanisms such as within the context of string theory. But there is a 
natural solution to these two problems. As discussed before we can
construct the same Dirac mass matrices in the models with the RGE
suppression factors (except for that of the neutrino which is
suppressed by the very large volume factor). In this alternative
construction, no gauge non-singlet matter lives in the bulk and
clearly the proton can be stable to all-orders in perturbation theory
if, for example, we impose the $Z_2$ parity symmetry under which the
dangerous baryon number violating fields have odd 
parity \cite{tev-gut}. Moreover, in the expression of the suppression
factor (\ref{gut-ep}), the number $n'$ is now replaced with that
determined by the dimension of each higher dimensional operator
including the bulk gauge-singlet field $\theta$. Compared to the
approach utilizing the volume factor $\epsilon$, the `unit'
suppression factor 
becomes $\epsilon' (\equiv \epsilon^{(n+2)/2})$. E.g., for $n=1$,
$\epsilon' \simeq (1/\sqrt{20})^{3/2} \simeq 0.1$, which is just the
required value for the observed hierarchies of fermion masses. In this
way all the problems stated above can be cured in this alternative
approach with the RGE effects.

In summary, we have demonstrated that in models with the (large) 
extra space-time dimensions, the realistic fermion mass hierarchy can
be naturally obtained by the suppression factors originated from the
existence of extra dimensions. In the models presented in this letter,
not only the hierarchical structure but also the large mixing angle in
the lepton sector can be realized. Of course, there are still very
wide possibilities with these extra dimensions, and it is interesting
to consider that the structure of the extra dimensions and fermion
mass hierarchy may be strongly related.

\subsection*{Acknowledgments}

The author would like to thank M.\ Bando and T.\ Noguchi for useful
discussions and careful reading of the manuscript. This work was
supported in part by the Grant-in-Aid for JSPS Research Fellowships.

\end{document}